\documentclass[twocolumn]{aastex63}

\usepackage[caption=false]{subfig}
\usepackage{floatrow}

\newcommand{\NEW}[1]{#1}

\received{\today}
\revised{}
\accepted{}

\shorttitle{Stick-slip dynamics in penetration experiments on simulated regolith}
\shortauthors{Featherstone et al.}

\graphicspath{{./}}

\begin{document}

\title{Stick-slip dynamics in penetration experiments on simulated regolith}


\correspondingauthor{Karen Daniels}
\email{kdaniel@ncsu.edu}

\author[0000-0001-8076-9669]{Jack Featherstone}
\affiliation{Department of Physics, North Carolina State University, Raleigh, NC, 27675, USA}

\author{Robert Bullard}
\affiliation{Department of Physics, North Carolina State University, Raleigh, NC, 27675, USA}

\author{Tristan Emm}
\affiliation{Department of Physics, North Carolina State University, Raleigh, NC, 27675, USA}

\author{Anna Jackson}
\affiliation{Department of Physics, North Carolina State University, Raleigh, NC, 27675, USA}

\author{Riley Reid}
\affiliation{Department of Physics, North Carolina State University, Raleigh, NC, 27675, USA}

\author{Sean Shefferman}
\affiliation{Florida Space Institute and Department of Physics, University of Central Florida, Orlando, FL, 32816, USA}

\author[0000-0001-5545-4454]{Adrienne Dove}
\affiliation{Florida Space Institute and Department of Physics, University of Central Florida, Orlando, FL, 32816, USA}

\author[0000-0001-8269-6408]{Joshua Colwell}
\affiliation{Florida Space Institute and Department of Physics, University of Central Florida, Orlando, FL, 32816, USA}

\author[0000-0002-3517-1139]{Jonathan E. Kollmer}
\affiliation{Department of Physics, North Carolina State University, Raleigh, NC, 27675, USA}
\affiliation{Universit\"at Duisburg-Essen, Lotharstr. 1, 47057 Duisburg, Germany}

\author[0000-0001-6852-3594]{Karen E. Daniels}
\affiliation{Department of Physics, North Carolina State University, Raleigh, NC, 27675, USA}

\begin{abstract}

The surfaces of many planetary bodies, including asteroids and small moons, are covered with dust to pebble-sized regolith held weakly to the surface by gravity and contact forces. Understanding the reaction of regolith to an external perturbation will allow for instruments, including sensors and anchoring mechanisms for use on such surfaces, to implement optimized design principles. We analyze the behavior of a flexible probe inserted into loose regolith simulant as a function of probe speed and ambient gravitational acceleration to explore the relevant dynamics.
The EMPANADA experiment (Ejecta-Minimizing Protocols for Applications Needing Anchoring or Digging on Asteroids) flew on several parabolic flights. It employs a classic granular physics technique, photoelasticity, to quantify the dynamics of a flexible probe during its insertion into a system of bi-disperse, cm-sized model grains. We identify the force-chain structure throughout the system during probe insertion at a variety of speeds and for four different levels of gravity: terrestrial, martian, lunar, and microgravity.
We identify discrete, stick-slip failure events that increase in frequency as a function of the gravitational acceleration. In microgravity environments, stick-slip behaviors are negligible, and we find that faster probe insertion can suppress stick-slip behaviors where they are present.
We conclude that the mechanical response of regolith on rubble pile asteroids is likely quite distinct from that found on larger planetary objects, \NEW{and scaling terrestrial experiments to microgravity conditions may not capture the full physical dynamics.}

\end{abstract}


\section{Introduction} \label{sec:intro}

\NEW{Understanding how regolith reacts to external disturbance is integral to developing safe and efficient techniques for sample collection and anchoring on extraterrestrial bodies. Beyond the terrestrial gravity that we experience every day -- and in particular, when gravity is weaker -- exploration and sampling missions must carefully consider their destination when designing instruments. This holds true for destinations like our moon \citep{nash_evaluation_1993} or other planets, where gravity is at least comparable to that of Earth, but also for the extreme case of asteroids, where gravity can be 4-5 orders of magnitude weaker \citep{hestroffer_small_2019}. As examples of this latter \textit{microgravity} regime, indeed the surfaces of both (101955) Bennu \citep{barnouin_shape_2019} and (162173) Ryugu \citep{watanabe_hayabusa2_2019} have been shown to be composed of loosely bound rubble.}
Many questions regarding the formation  \citep{cheng_reconstructing_2020, michel_collisional_2020}, surface features \citep{shinbrot_size_2017, bogdan_laboratory_2020}, and mechanical stability \citep{sanchez_cohesive_2020, scheeres_scaling_2010} of these asteroids still remain open.

There have been a number of numerical simulations \citep{maurel_numerical_2018, thuillet_numerical_2018} as well as laboratory experiments \citep{colwell_low-velocity_1999, colwell_ejecta_2008, altshuler_settling_2014, li_3d_2016, brisset_regolith_2018, brisset_regolith_2020} on low-velocity impacts onto simulated regolith surfaces under low gravity conditions. These impacts are not only relevant for understanding the formation of these bodies, but can also give insight into the inner composition of an asteroid \citep{quillen_impact_2019}. For more detailed exploration, however, sampling the surface is necessary.

The recent sample collection by NASA's OSIRIS-REx mission is an example of how instrumentation methods depend on the details of granular-structure interactions. The Touch-And-Go Sample Acquisition Mechanism (TAGSAM)  involved expelling pressurized nitrogen gas to dislodge regolith from the surface, which was then captured within the sampling head \citep{bierhaus_osiris-rex_2018}. The sample collection was successful, though an extensive area around the sampling site was disturbed \citep{dunn_nasa_2021}, and an over-collection of regolith occurred \citep{bierhaus_bennu_2021}. Hayabusa2's instruments take a similar approach to those of the OSIRIS-REx mission, capturing ejected regolith after disturbing the surface of Ryugu \citep{sawada_hayabusa2_2017}. In this case, the disturbance is caused by a projectile, but the underlying dynamics still depend sensitively on the details of the granular material, \NEW{like the particle size \citep{li_3d_2016}. The magnitude of the surface gravity is another property that has a significant effect on the dynamics; an understanding of low-gravity granular interactions would thus aid in the design of future regolith collection missions.}

\begin{figure*}
    \centering
    \includegraphics[width=.98\linewidth]{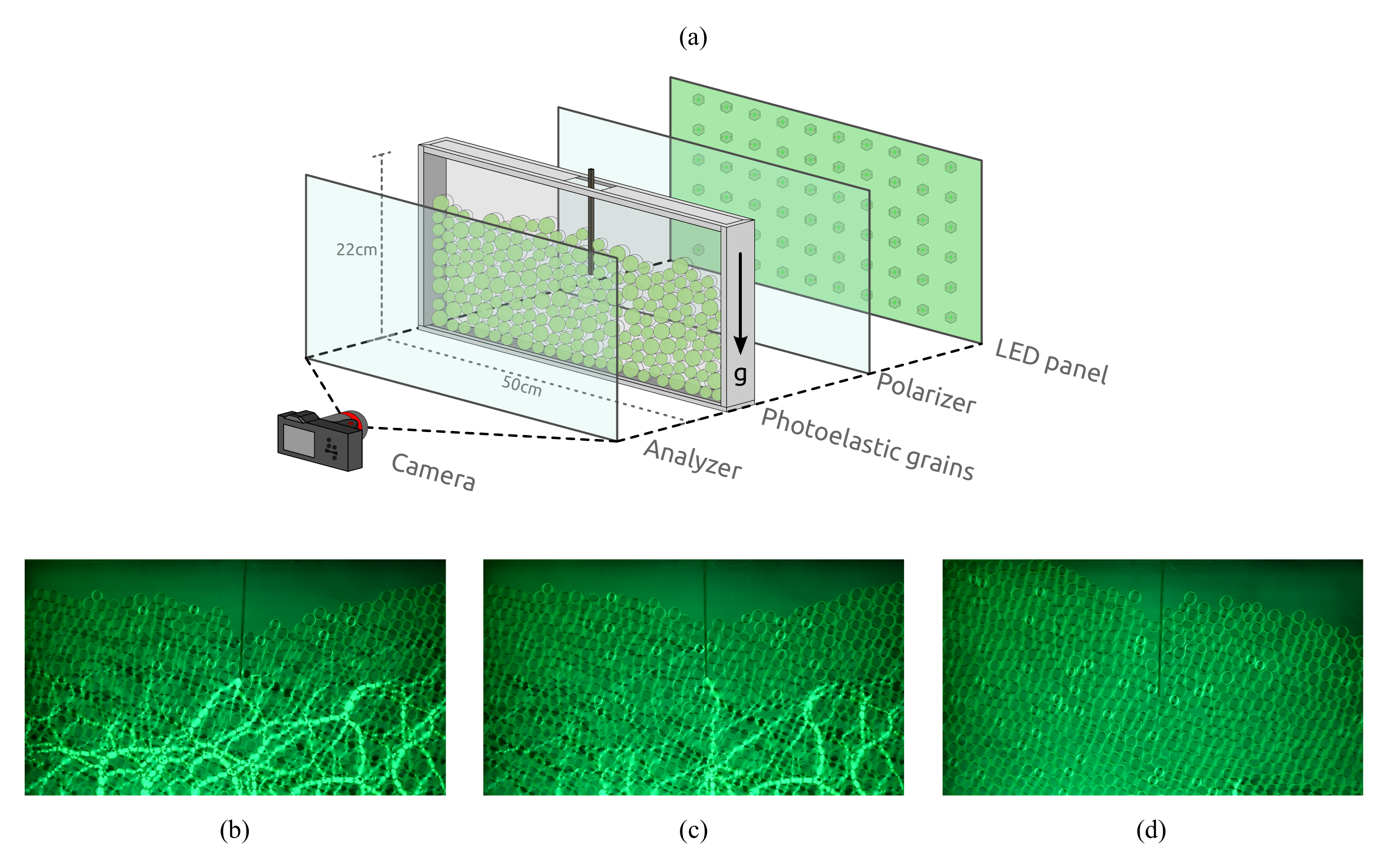}

    \caption{{\bf (a)} \NEW{Schematic of the EMPANADA-2D apparatus, showing the photoelastic particles, camera, and optical polarizers used to visualize forces.} {\it Bottom:} Qualitative comparison of data from {\bf (b)} martian, {\bf (c)} lunar, and {\bf (d)} microgravity trials. Brighter particles are subject to more intense interparticle stress.}
    \label{fig:Overview}
\end{figure*}

In terms of mechanically manipulating and collecting a sample, there are presently no established safe and effective protocols to guide the interaction of space probes with low-gravity, poorly-consolidated material \citep{wilkinson_granular_2005}. 
\NEW{The InSight mission \citep{lichtenheldt_soil_2016}, is currently investigating seismological activity on Mars, using a rigid, self-propelled penetrator (mole) to measure regolith thermal conductivity and thermal gradients. The mole's pile-driver design \citep{komle_pile_driving_2015} was built to be hammered into a surface using a mechanism housed within the mole itself, with an aim to penetrate up to 5~m deep. This design has been found to be very effective at penetrating regolith in terrestrial environments \citep{wippermann_penetration_2020}, but was not nearly as successful on Mars \citep{good_nasa_2021}, highlighting the sensitivity of the dynamics to in-situ grain properties. When particle-ejecta can easily pose concerns for probe or craft safety -- as in the case of asteroids -- we are confronted with an even narrower margin of error during operation.}

For rubble-pile asteroids, where the escape velocities can be on the order of cm/s \citep{sanchez_strength_2014,daniels_rubble-pile_2013}, the development of these ejecta-minimizing operations will be particularly important.
One promising technique, inspired by the growth of plant roots \citep{kolb_radial_2012, wendell_experimental_2012, fakih_root_2019}, is the use of slow \NEW{($\sim$~mm/s)} flexible probes \citep{kollmer_digging_2016}. \NEW{In particular, \citet{fakih_root_2019} find, through modeling roots as variably flexible intruders, that disturbance to a heterogeneous granular material is lessened as flexibility increases.} The dynamics of such a flexible, slender probe interacting with a granular material in low gravity are unexplored, \NEW{and may lead to useful sampling techniques. Furthermore, the similarities and differences between probe-insertion and impact-dynamics have not been extensively mapped.}

\NEW{One way to characterize intruder disturbance within a granular material is to examine the spatial distribution of interparticle forces, commonly known as force chains \citep{pica_ciamarra_dynamics_2004, clark_particle_2012, zheng_intruder_2018, bester_dynamics_2019}. Granular materials often transmit stress anisotropically \citep{majmudar_contact_2005}, with some particles bearing significantly larger loads than others. This force chain structure, and the subsequent failure patterns \citep{liu_spongelike_2021}, depend on the material loading history \citep{kollmer_betweenness_2019}. 

As in \citet{clark_particle_2012} and \citet{zheng_intruder_2018}, we employ photoelasticity to experimentally investigate the force chain structure of an intruder, applied to astrophysical contexts. Photoelastic materials have stress-dependent polarizations, which means that an appropriate configuration of polarizing filters allows for interparticle forces to be imaged directly \citep{daniels_photoelastic_2017, abed_zadeh_enlightening_2019}.
}

In this paper, we present investigations of a prototypical surface interaction: the insertion of a flexible probe into a granular material under various levels of gravity. This project is named EMPANADA (Ejecta-Minimizing Protocols for Applications Needing Anchoring or Digging on Asteroids), and is inspired by previous work on the PRIME project \citep{brisset_regolith_2018}. In these new experiments, we slowly insert a flexible intruder into a vertical bed of round, quasi-two-dimensional elastic grains.

\NEW{Concurrent probe insertion experiments were performed with a three-dimensional system, more similar to the PRIME project, reported in more detail by \citet{kollmer_probing_2021}. We find that the two-dimensional system offers several advantages in terms of data interpretation.
Of particular importance is the ability to track particles within the simulated regolith, according to both their positions and their interparticle forces. Here, we  focus our analyses on the intermittent dynamics of the two-dimensional heterogeneous force chains; preliminary analysis of the three-dimensional system (via other methods) shows similar trends. The primary mode of energy dissipation --- via friction and inelasticity between adjacent particles --- is fundamentally the same in two- or three-dimensions, although the three-dimensional particles will have more dissipation as a result of higher contact numbers.}

We find that both the magnitude and frequency of rearrangements increase as a function of the gravitational acceleration. For experiments at terrestrial, martian, and lunar gravity, faster insertion speeds suppress \NEW{stick-slip \citep{albert_stick-slip_2001, braun_nanotribology_2006}} dynamics within the granular bed. That is, the number of discrete stick-slip failure events is reduced, without a significant increase in the intensity of each individual event; this is surprising because a faster probe injects more kinetic energy. For the microgravity case, we are not able to identify any specific relationship between insertion speed and granular dynamics; in this regime, the \NEW{fluid-like} behavior of the grains is not well described by discrete stick-slip events.

\section{Methods} \label{sec:methods}

We performed experiments on the apparatus described in Fig. \ref{fig:Overview}a, which was both flown on parabolic flights and used in laboratory experiments on earth's surface. The setup centers around a backlit, vertical bed of \NEW{bi-disperse photoelastic disks}, and a flexible probe mounted on a motor driven slider. \NEW{The particles are individually cut from flat sheets of Vishay PSM-4 according to the methods described in \citet{daniels_photoelastic_2017}, with diameters of $0.9$cm and $1.1$cm. The enclosure dimensions of approximately $50$~cm wide by $22$~cm tall were chosen to fit within the existing PRIME enclosure, while also being large enough to minimize lateral boundary effects \citep{desmond_random_2009}. The camera captured images of the light passing through the particles and two optical polarizers required for photoelastic visualization (\textit{polarizer} and \textit{analyzer}).} Sample images from this setup in three different levels of gravity -- martian, lunar, and micro -- are shown in Figs. \ref{fig:Overview}b, \ref{fig:Overview}c, and \ref{fig:Overview}d.

\NEW{During each trial, a flexible rectangular probe made from acrylic, with a $3\times1.5$~mm cross-section, is driven approximately $10$~cm into the granular system.} The probe is flexible enough to navigate through the grains by either bending itself or slipping past individual particles, both of which can cause varying levels of disturbance in the system. The probe is attached to the stepper motor \NEW{(MakerBot 50002)} via a load cell \NEW{(CLZ616C)}, which is driven into the medium at a constant speed. This load cell records forces parallel to the probe as it traverses the system \NEW{at a rate of $40$~Hz, and a camera records videos of the process at $30$~Hz}.
We conducted experiments -- consisting of the probe being driven into the granular material -- under four different gravity conditions: terrestrial, martian, lunar, and microgravity. The latter three experiments were conducted during a parabolic flight campaign (Zero-G Corporation); terrestrial trials were completed on the same apparatus after its return. 

\begin{figure}
    \centering
    
    \includegraphics[width=.95\linewidth]{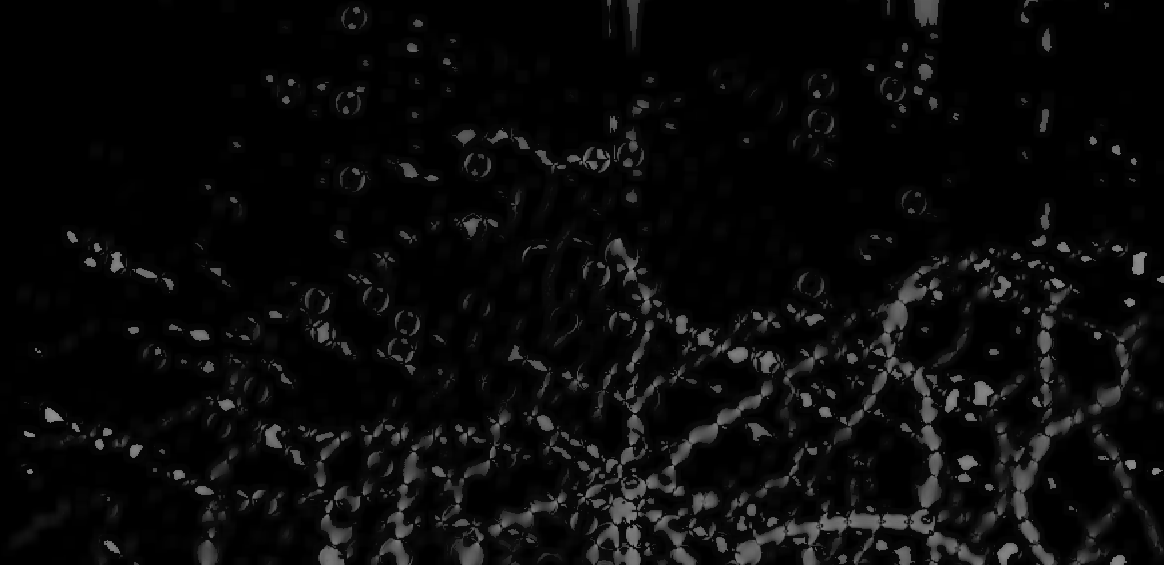}
    \label{fig:AfterEditSample}
    
    \caption{\NEW{Sample image of the image processing applied to each image, based on the original shown in Fig. \ref{fig:Overview}c. Extreme pixels with brightness values $I \in [0, 255]$ are exponentially enhanced, $\tilde{I} = \exp[|I - 127 |/127]$, focusing attention on the portions of the image that represent interparticle forces.}}
    \label{fig:ProcessingComparison}
\end{figure}

The photoelastic particles, having stress-dependent polarization, indicate the strength of local, inter-particle forces via their brightness \citep{daniels_photoelastic_2017, abed_zadeh_enlightening_2019}. Video of the granular bed allows for observation of the brightness in each frame throughout the insertion process, giving a profile of the net forces in the system. This metric can then be modified and extended to highlight the general behavior of the insertion process, including identifying discrete stick-slip events. Due to safety considerations, the particle enclosure is constructed from acrylic sheets; these act as fractional waveplates, and slightly rotate the polarization of the light transmitted through the particles. This effect manifests as some force chains appearing dark in the video frames (eg. Fig. \ref{fig:Overview}c); we adjusted for this by preprocessing the images to highlight the force chains, regardless of their orientation in the raw images (Fig. \ref{fig:AfterEditSample}). 
\NEW{This is done by transforming the raw data, $I$, to emphasize those pixels which are much dimmer/brighter than the typical value; this suppresses the contributions from unimportant features such as particle outlines. In addition, we correct for the vertical linear light gradient, to account for the uneven lighting conditions on the parabolic flight. We denote this new quantity as $\tilde I$.}

\begin{figure}
    \centering
    \includegraphics[width=.98\linewidth]{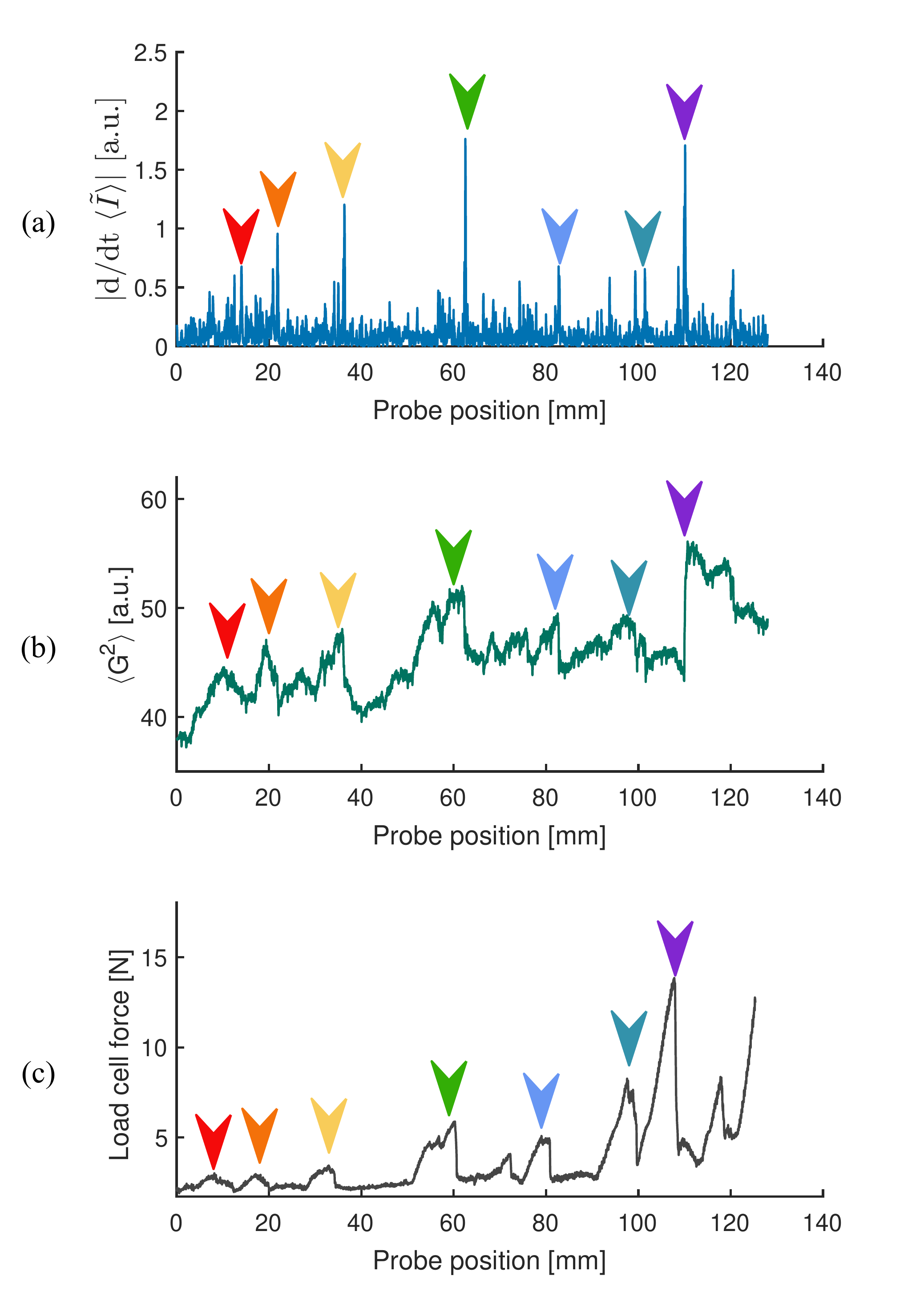}

    \caption{ Comparison between the granular dynamics as measured by {\bf (a)} the time derivative of \NEW{average $\tilde I$ (transformed brightness), {\bf (b)} average $G^2$ (local gradient, squared, of the image brightness)}, and {\bf (c)} the load cell readings on the probe. Arrow colors indicate the same event across all three graphs. Units on the first two plots cannot be directly converted to physical units, and therefore are left in arbitrary units (a.u.). }
    \label{fig:SampleComparison}
\end{figure}

\begin{figure}
    \centering
    \includegraphics[width=.98\linewidth]{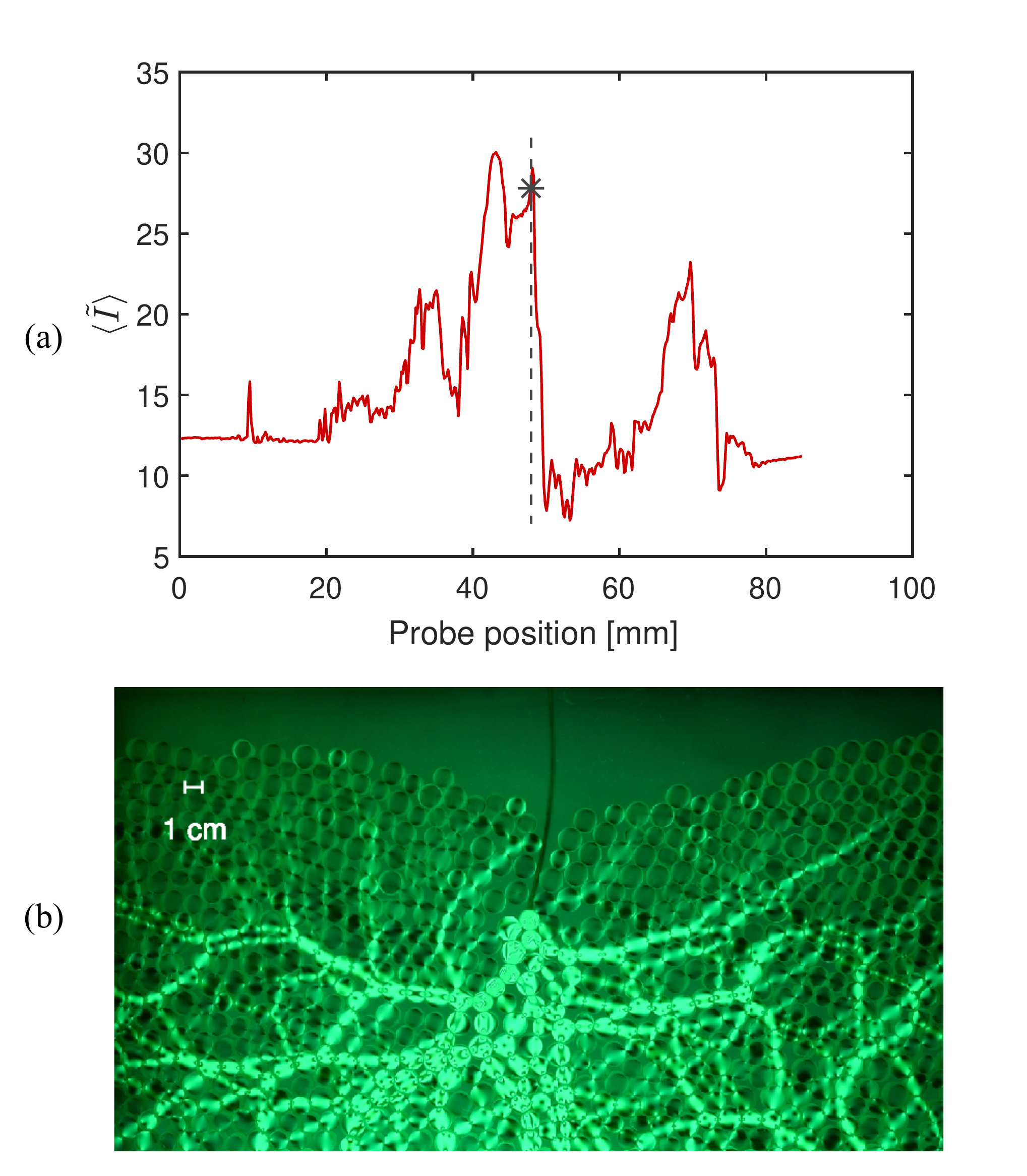}
    
    \caption{{\bf (a)} The plot of \NEW{transformed frame brightness, $\tilde I$,} for a martian trial. Annotation on the plot represents the time that corresponds to the frame shown in {\bf (b)}. The grains are made of photoelastic materials, meaning the regions under higher stress appear brighter. \NEW{The impending failure in the bottom raw photoelastic image can be seen in the plot of $\tilde I$ as the sharp decline immediately following the annotation.}}
    \label{fig:RealTimeSample}
\end{figure}

\NEW{Although the load cell is able to directly record the forces on the probe, we primarily analyze the photoelastic response, as calibrated load cell data is only available for terrestrial runs due to a malfunction that occurred during the parabolic flights.} A comparison between the direct measurement and this image analysis (Figs. \ref{fig:SampleComparison}a, \ref{fig:SampleComparison}b, \ref{fig:SampleComparison}c) demonstrates that measuring the brightness encapsulates the information present from the load cell, along with additional information \NEW{that could represent} forces perpendicular to the probe not contained in the former metric.  While granular physics experiments often use the local gradient of image brightness (the $G^2$ method, \citet{abed_zadeh_enlightening_2019}) to quantify forces instead of the average brightness, we find that the brightness was most reliable for counting the number of discrete events and have therefore used that measure in our analyses.

To gather data in lower gravity conditions, the apparatus was flown on a parabolic flight campaign (Zero-G Corporation), yielding 41 total trials, each corresponding to martian (6), lunar (5), or microgravity (30). The probe insertion process was run at several different speeds for each level of gravity over the course of two flights. Although some $g$-jitter was present during these flights, the majority of the trials were suitable for semi-quantitative analysis. Those which we identified to have major deviations from the expected conditions -- e.g. net acceleration pointing upwards during certain microgravity trials -- were excluded from our analyses. After the flight campaign, the apparatus was brought back to the lab, and additional data was taken in terrestrial gravity; these trials were able to cover a more extensive range of speeds. The final data set includes trials run at three ``planetary'' gravities -- lunar, martian, and earth -- as well as asteroid-like microgravity.

A sample frame from a trial can be seen in Fig. \ref{fig:RealTimeSample}b, with the corresponding brightness plot shown above (Fig. \ref{fig:RealTimeSample}a), where the image time is indicated by the dashed line. The force chains can be identified as bright paths through the grains. As can be seen in the video frame, the probe is currently exerting a force on a single grain, which then propagates through the rest of the system. Once the force becomes strong enough, the probe will slip off the grain it is in contact with, causing nearby grains to reconfigure. This can identified in the brightness plot, represented as the sharp decline immediately after the current time marked on the figure.

\NEW{Motivated by this behavior, we next look at the time derivative of $\tilde I$ as a metric for how the medium is changing throughout the insertion process, calculated using a central difference stencil.}
\NEW{We identify each local maximum in this discretized time derivative of $\tilde I$ as a \textit{stick-slip event} \citep{albert_stick-slip_2001, daniels_force_2008}, conditional on being sufficiently separated from other local maxima by at least the time $t_c = 0.5$~s. The event-counting algorithm, and subsequently the trends presented below, are largely insensitive to reasonable choices of the threshold, separation time, and brightness rescaling.}

\begin{figure}
    \centering
    \includegraphics[width=.98\linewidth]{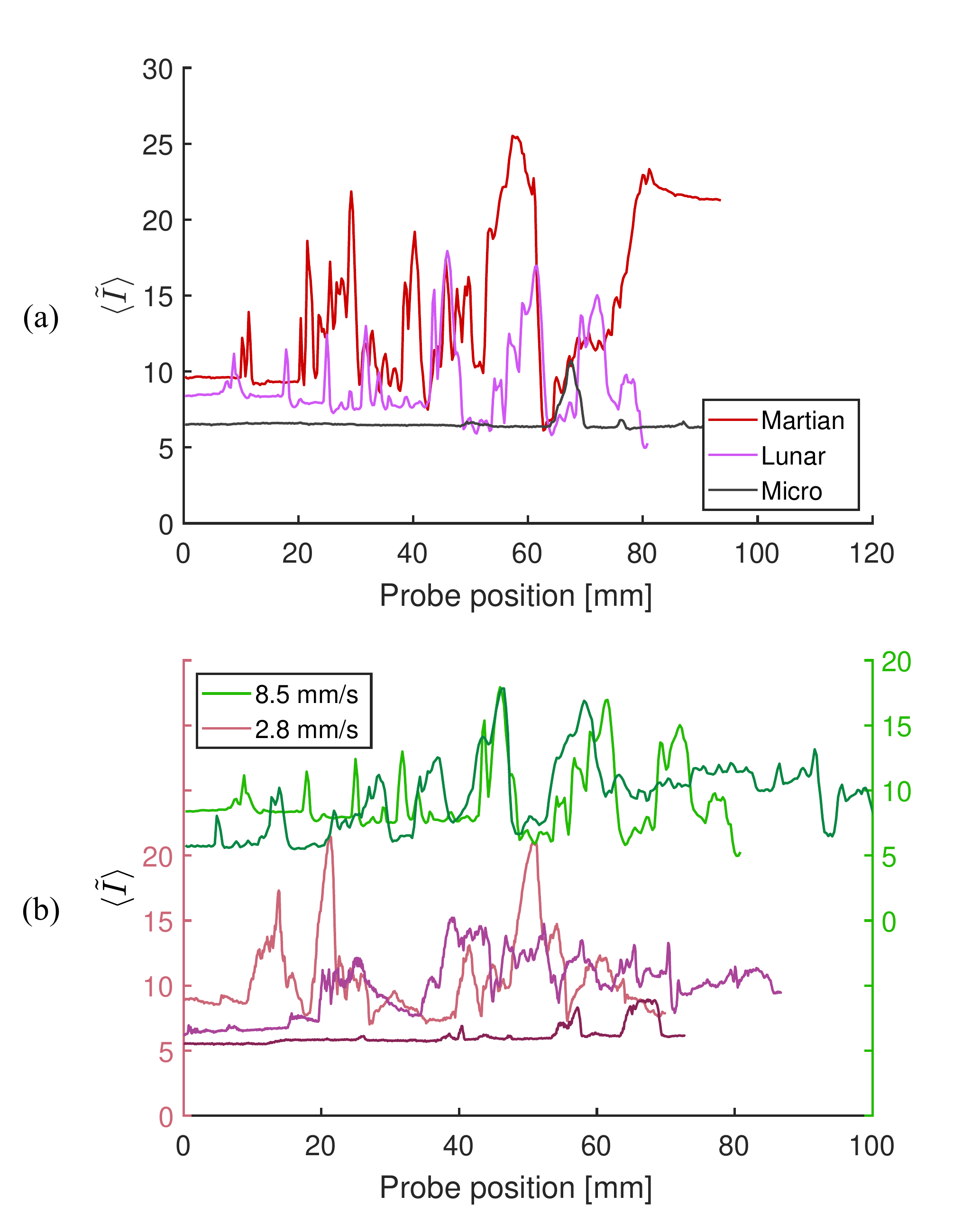}

    \caption{{\bf (a)} The average brightness of each video frame for a trial in each level of gravity simulated during the parabolic flight campaign, all with probe insertion speed of 8.5 mm/s. {\bf (b)}  The brightness profiles for several trials conducted in lunar gravity, with multiple trials at the same insertion speed grouped together. \NEW{Units for average image brightness, $\tilde I$ (see  Fig.~\ref{fig:ProcessingComparison} caption), are rescaled pixel intensity from camera, which act as a proxy for average interparticle force}.}
    \label{fig:SampleAndSpeed}
\end{figure}

\section{Results} \label{sec:results}

\begin{figure*}
    \centering
    \includegraphics[width=.98\linewidth]{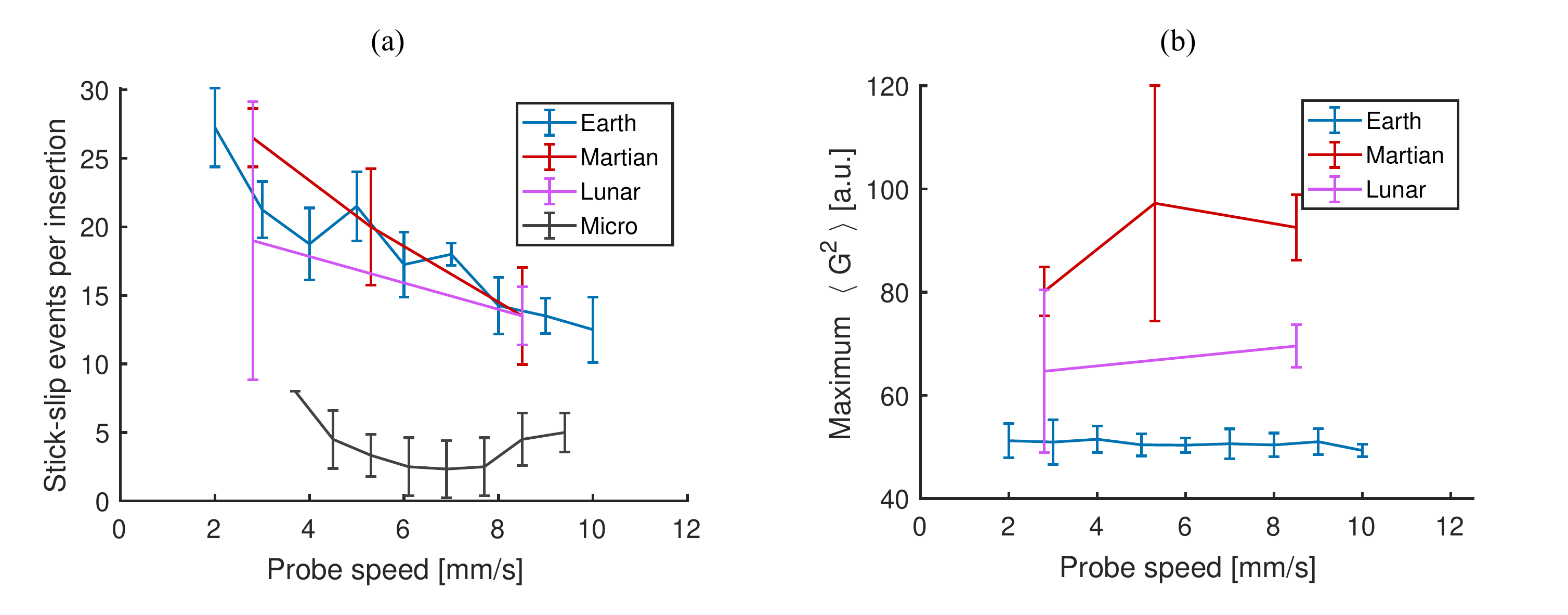}
    
    \caption{{\bf (a)} The discrete number of stick-slip events for each $5$cm insertion, counted as local maxima in the discretized time derivative of the average frame brightness. Appropriate statistical error bars are included for gravity/speed pairs that had more than one data point. {\bf (b)} A comparison of the maximum local gradient squared during the terrestrial, martian, and lunar trials, taken as a proxy for the maximum instantaneous force. Each connected set of points (same gravity) were taken under the same lighting conditions, but cannot be compared between sets.}
    \label{fig:SlipAndLoad}
\end{figure*}

Figure \ref{fig:SampleAndSpeed}a shows a comparison of the photoelastic response (representing the total interparticle forces in the system) during probe insertion for 3 different trials, covering the three extraterrestrial levels of gravity that were tested -- martian, lunar, and microgravity. Qualitatively, higher gravitational acceleration gives a more sharply peaked average brightness, corresponding to stronger forces throughout the granular bed. Martian and lunar gravity show easily identifiable stick-slip events in the form of local maxima in the time dependent intensity data, confirmed by visual inspection of the corresponding videos. The insertion dynamics in microgravity exhibit more subtle behavior: there are features of the brightness curve that may be described as discrete events, but they don't seem to represent the dominant pathway for spatial reconfiguration. Per this behavior, we can show (Fig. \ref{fig:SlipAndLoad}a) that attempting to describe a microgravity system as a sequence of stick-slip events -- as would be done for the other trials -- would not be appropriate.

Unlike the comparison between different levels of gravity, different insertion speeds don't present an immediately recognizable trend in the corresponding granular dynamics. This can be seen in Fig. \ref{fig:SampleAndSpeed}b, where a comparison of different speeds is shown for five lunar gravity trials. On average, the trials at lower speeds display no obvious differences from the ones conducted at higher speeds; a more quantitative analysis including additional trials is required to identify any trends. This is representative of the other three levels of gravity, as similar comparisons for these cases also showed no significant qualitative differences.

Figure \ref{fig:SlipAndLoad}a displays the number of stick-slip events per probe insertion evaluated for all of the levels of gravity and all insertion speeds. For the planetary levels of gravity (all but microgravity), there is a clear downward trend in the number of discrete stick-slip events for higher speeds, which is reasonably within the statistically-calculated error bars. This demonstrates that for the insertion protocol used here (constant driving velocity), a higher insertion speed suppresses stick-slip behavior in conditions with significant gravity. For the microgravity case, there is not a strong trend to be seen, confirming our qualitative predictions that discrete rearrangements do not properly characterize the granular dynamics in this case; instead, the behavior is more fluid-like.

We next turn our attention to the intensity associated with each of these events. We evaluated the maximum frame-averaged $G^2$ value for each video -- representing the maximum magnitude of instantaneous force -- which is shown in Fig. \ref{fig:SlipAndLoad}b. Microgravity has been excluded from this comparison since it is not well-described by discrete events. No significant increase or decrease can be seen for any of the trials when considering the statistically-determined error bars. This demonstrates that the frequency of events is more strongly a function of the insertion speed than the intensity, although more data would be needed to establish this quantitatively. 

\section{Discussion} \label{sec:disc}

Considering the frequency of stick-slip events (Fig. \ref{fig:SlipAndLoad}a) and their intensities (Fig. \ref{fig:SlipAndLoad}b) together suggests that the decrease in frequency at higher speeds is the sole effect of varying the insertion speed. \NEW{This decrease of stick-slip behavior likely represents an overall decrease in the disturbance caused by the intruder} in the granular material for situations in which significant gravity is present. For these cases, the acceleration due to gravity is strong enough that looking at discrete failure events captures a significant portion of the granular dynamics. 

Taking a broader perspective, this disturbance minimization must be limited by two constraints. First, the flexibility of the probe at some point will transition from plowing through the material to bending \citep{algarra_bending_2018}. The other limit is the inertia of the particles; for interactions faster than the speed of sound in the material, one would expect to produce a shock \citep{hancock_simulating_2020}. This is especially relevant considering that the speed of sound in a granular material is a function of confining pressure, and therefore ambient gravity \citep{gomez_uniform_2012}. We propose that an optimum insertion velocity must exist, for which disturbances of the granular bed are minimized, and that further experiments are needed to map out these dependencies. \NEW{This is in contrast to the case of an impact on a regolith surface, for which the volume of ejected regolith seems to only increase with the impactor's kinetic energy \citep{colwell_ejecta_2008}.}

\NEW{We expect that the power required to drive a flexible probe into a granular material is significantly less than the equivalent power required for a pile-driver design. While a pile-driver can create shocks in the material \citep{gomez_uniform_2012}, driving a flexible intruder the comparatively-low speeds considered in this study will not. From the force data gathered in terrestrial gravity (Fig. \ref{fig:SampleComparison}c) and the known insertion speeds, we find that this power to drive the probe is on the order of milliwatts for our system.}

For the microgravity trials, a dependence on insertion velocity was not observed, as exemplified by the rather flat curve in Fig. \ref{fig:SlipAndLoad}a. This is also qualitatively supported by the lack of extended force chains present in the microgravity images (Fig. \ref{fig:Overview}d). \NEW{There likely exists a phase transition between the lunar- and micro- gravity datapoints, representing this change from solid-like to fluid-like behavior. Within this} fluid-like regime, a different metric would need to be identified for characterizing the granular dynamics. \NEW{The fluid-like and solid-like behaviors are sufficiently distinct that simply scaling terrestrial experiments to microgravity conditions may not be sufficient to robustly predict the relevant dynamics.}

\NEW{Because the microgravity regime is not described well by looking at stick-slip failure, it is necessary to consider an alternative description of the granular dynamics, possibly including other avenues for rearrangement like creeping behavior.} The dynamics of inserting a flexible intruder into a granular material is also a current topic of interest within the granular physics community \citep{algarra_bending_2018, mojdehi_buckling_2016}, from which additional methods can be drawn. While there are similarities between the terrestrial and microgravity cases, \NEW{our study suggests that experiments} in terrestrial gravity cannot reasonably be scaled to describe the microgravity case. \NEW{This result is consistent across both the 2D and 3D experiments conducted in association with this investigation; while we have focused primarily on the 2D case here, \citet{kollmer_probing_2021} outlines the commonalities between the two experiments. These both direct us to believe that the singular nature of the microgravity case necessitates in-situ experiments to properly quantify the dynamics.}

\section{Conclusions} \label{sec:concl}

\NEW{We have used photoelastic techniques borrowed from earth-based granular physics, not previously applied to astrophysical contexts, to understand regolith dynamics on planetary bodies. For cases in which gravity is on the order of terrestrial gravity, regolith behavior is well described by looking at force-chain structure and stick-slip dynamics. Importantly, investigating the heterogeneous force chains within a granular system in a variable gravity environment (including whether or not they are present) can provide information about how that system will react to external influence.}

\NEW{We make note of a transition between solid-like and fluid-like behavior with decreasing gravity, as well as a reduction of stick-slip dynamics where they are present with increasing probe insertion speed. Our data indicates that the solid-like and fluid-like behaviors are separated by a phase boundary located between our micro- and lunar- gravity datapoints: $.001 g_{\text{earth}}$ and $.166 g_{\text{earth}}$, respectively. A similar transition was observed in other recent investigations of granular phenomenon as a function of gravity, including confined dense flows \citep{shaebani_gravity_2021} and angle of repose \citep{elekes_expression_2021}. In \citet{shaebani_gravity_2021}, the transition zone lies in the range $0.1 g_{\text{earth}} \lesssim g \lesssim g_{\text{earth}}$.}

\NEW{We conclude that the largely unexplored flexible probe paradigm can be of great use for sampling and anchoring applications on planets and asteroids. This design demonstrates rich dynamics both as a function of gravity and insertion speed, which are of great interest in both astrophysical and terrestrial contexts.}

\acknowledgements This work was funded by NASA Grant number 80NSSC18K0269 (REDDI) and National Science Foundation grant number DMR-2104986, along with with undergraduate student funding from both the NC State Office of Undergraduate Research and the NC State Provost's Professional Experience Program. JK acknowledges funding by the German Aerospace Center DLR with funds provided by the Federal Ministry for Economic Affairs and Energy (BMWi) via Grant Number 50WM1943.

\bibliography{EMPANADA}
\bibliographystyle{aasjournal}

\end{document}